\documentclass{article}
\usepackage{fullpage}
\usepackage{amsfonts,amssymb}
\usepackage{amsthm}
\usepackage{tikz}
\usetikzlibrary{calc}
\usetikzlibrary{arrows}
\usepackage{tikz-3dplot}
\usepackage{color} 
\usepackage[fleqn]{amsmath}
\usepackage{amsthm}
\usepackage{amssymb}
\usepackage{amsfonts}
\usepackage{bbm}
\usepackage{epsfig}
\usepackage{times}
\usepackage{hyperref}
\usepackage{bm}
\usepackage{float} 
\usepackage{appendix} 
\usepackage{lscape}
\usepackage{cite}
\usepackage{mathrsfs}
\usepackage{appendix}
\usepackage{setspace}
\usepackage{mathtools}
\usepackage{authblk}
 
\theoremstyle{definition}

\begin{document}
	
		\title{Laws of Nature as Constraints}
		
		\author[1]{Emily Adlam} 
		\affil[1]{University of Western Ontario}
	
	\maketitle

\section{Introduction}

The laws of nature have come a long way since the time of Newton: quantum mechanics and relativity have given us good reasons to take seriously the possibility of laws which may be non-local, atemporal, `all-at-once,' retrocausal, or in some other way not well-suited to the standard  dynamical time evolution paradigm.  But many extant realist approaches to lawhood are based on the paradigm of Newtonian laws, and thus they face significant challenges when we try to apply them to laws outside the time evolution picture. In light of these developments, we are in need of a new  approach to lawhood which is better suited for laws outside the time evolution paradigm.  In the present paper, we argue that this can be achieved within an approach which conceptualizes lawhood in terms of modal structure, and we set out a general framework in which the modal structure associated with a  wide class of possible laws can be expressed and studied. 

This project is motivated in large part by the recognition that the ideas we have about what laws look like inevitably shape the types of laws which scientists formulate. Thus the limitations of our current notion of lawhood translate to limitations on scientific thought:  although laws outside the dynamical time-evolution paradigm are gaining increasing prominence in physics, many physicists are reluctant to accept them and there remains confusion about their conceptual significance. Moreover, the development of these sorts of approaches is currently hampered by attempts to assimilate them to a time-evolution picture - for example Kastner, observing that modern-day proponents of retrocausal models often attempt to express them in dynamical terms, comments that `\emph{the dynamical stories presented in these interpretations are inconsistent with the ontology they require.}'\cite{doi:10.1063/1.4982766} So in order to facilitate further progress within these research programmes it is crucial to develop a conceptually clear way of thinking about lawhood which straightforwardly accommodates laws outside the time evolution paradigm. The framework presented in this article is intended to help emerging research on non-dynamical laws propagate into broader scientific thought and to give scientists better conceptual resources for their explorations of various kinds of non-standard laws. 

Moreover, there are many philosophical topics where the notion of lawhood plays a crucial part: for example, laws are often invoked in the analysis of causation\cite{sep-causation-counterfactual}, explanation\cite{hempel1965aspects}, determinism\cite{Butterfielddeterminism}, scientific confirmation\cite{Goddard1977-GODTPO}, prediction\cite{Dorst2019-DORTAB-2} and free will\cite{Hobart1934-HOBFWA}. But these analyses frequently assume, either explicitly or implicitly, that the laws in question take the standard Newtonian time-evolution form, and thus in the light of developments in modern physics which indicate that laws may not always take this specific form, there's a clear need for us to re-evaluate the role that laws play in these applications. Furthermore, in many of these applications it makes a difference whether the account of lawhood is realist or non-realist - for example, one might well draw different conclusions about determinism and its relationship with free will in the realist setting where laws \emph{make} things happen versus in the Humean best-systems setting where laws are simply read off whatever does happen. So in order to make progress on these topics we are in need of a realist way of thinking and talking about laws which accommodates various sorts of laws outside the time-evolution paradigm: the framework presented in this article is intended as a first step towards addressing these questions.

At this juncture, we should clarify that this paper is focused  on the fundamental laws of physics - Schrodinger's equation, the Dirac equation, the Einstein equations and their ilk, as well their as yet unknown successors. We have no doubt that interesting issues may arise in attempting to expand the framework suggested here to the laws of the special sciences and ceteris paribus laws, but that is a task for another day.

\section{Physics Beyond Time Evolution \label{intro}}

Newton bequeathed to us a picture of physics in which the fundamental role of laws is to give rise to time evolution: the Newtonian universe can be regarded as something like a computer which takes in an initial state and evolves it forward in time\cite{Wharton}. Newton's vision has been enormously influential on the philosophy of science and lawhood, and even today many philosophical discussions about lawhood use Newton's laws as an example, presumably on the grounds that $F = ma$ is a simple case which is presumed to share most of the relevant features of more complex laws. 

But there have been several major conceptual revolutions in physics since the time of Newon, and thus we should not necessarily expect that accounts of lawhood based on a Newtonian time-evolution picture will be well-suited to the realities of modern physics. In this section, we will discuss several important examples of laws appearing in modern physics which look very different to the traditional Newtonian time evolution laws. This will provide motivation for the claim that a more general realist framework is needed, and also  provide some inspiration for the overall structure of that framework. Note that it is not my intention to argue that all the putative laws mentioned in this section are definitely fundamental laws of physics, or indeed that these examples prove beyond doubt that we must abandon the time-evolution picture. The point is merely that all of these types of laws are taken seriously by at least some sectors of the scientific community, and thus if we accept the naturalistic notion that our account of lawhood should be led  by scientific practice rather than pre-existing metaphysical commitments, we need an account of lawhood which is capable of accommodating the same range of possibilities as science itself.

\subsection{Lagrangian approaches \label{Lagrange}} 

We don't even need to go beyond classical mechanics to encounter laws which don't work well within the standard time-evolution picture. For although it is most common to conceptualise classical mechanics in terms of the Newtonian schema\cite{Smolinref} in which laws act on states to produce time evolution, there is also an alternative Lagrangian description of classical mechanics in which systems are required to take the path which optimizes a quantity known as the Lagrangian\cite{brizard2008introduction}. Lagrangian methods are well-recognised as valuable mathematical tools, but they have not usually been taken seriously as possible descriptions of reality, presumably because the Lagrangian is optimized over an entire history and thus the Lagrangian description can't straightforwardly be understood within the standard time-evolution picture\cite{Adlamspooky, Wharton}. 

However path integrals - the analogue of the Lagrangian method within quantum mechanics\cite{feynman2010quantum} - have become so important to quantum field theory that increasingly we are seeing calls to reconsider the idea that the Lagrangian description is purely a mathematical trick\cite{hartlespacetime, Sorkinpath}. And indeed, there seems no reason other than a prejudice in favour of the time evolution picture to insist that the real laws must take a Newtonian form. Moreover, as argued by Wharton\cite{Wharton},  taking Lagrangian methods seriously leads to a novel `all-at-once' approach to lawhood in which we think of laws applying externally and atemporally to the whole of history at once\cite{Adlamspooky}.  Of course, in an `all-at-once' picture there is by construction no time evolution, so if we are to make sense of all-at-once laws of this form we are in need of an approach to lawhood which is not predicated on the time evolution picture.

\subsection{Non-dynamical approaches} 

There is a practice within physics of presenting theories in terms of their `kinematics' (state space) and `dynamics' (laws of evolution)\cite{Wigner995}, and thus it is common for philosophical accounts of laws of nature to conceptualise laws in this way. But there are a number of theories in modern physics where it doesn't make a great deal of sense to have a rigid division between `state space' and `evolution laws.' For example, the solution to the Einstein equations of General Relativity is not a state at a time but an entire history of a universe\cite{wald2010general}, so it doesn't seem to require any concept of time evolutionl. A time-evolution formulation of the Einstein equations does exist\cite{ringstrom2009cauchy, foures1952theoreme}, but the original global formulation remains central to research in the field and there seems no obvious reason to think that the time-evolution formulation must be more fundamental. Similarly, a number of approaches to quantum gravity, including quantum loop gravity\cite{Rovelli2008}, postulate laws which apply `all-at-once' to the whole of history,  rather than taking a state at a time and evolving it forwards. 

Note that the terms `kinematics' and `dynamics' are still sometimes used in the context of General Relativity and other theories without a notion of time evolution, where `kinematically possible' now refers to the set of models which are considered to be of the right mathematical type to represent a physically possible world, and `dynamically possible' refers to the set of models which do in fact represent a physically possible world\cite{pittphilsci16622}. But as Spekkens points out, there may be many different ways of splitting a theory into `kinematics' and `dynamics' and moreoever, the kinematics and dynamics of a theory can only ever be tested together, so there is no hope of empirically distinguishing which distinction between kinematics and dynamics is the correct one. Spekkens argues that we should therefore refrain from making this distinction altogether and instead think in terms of causal structure\cite{2012arXiv1209.0023S} (or perhaps \emph{modal} structure for greater generality). This line of argument suggests that simply decoupling the notion of dynamics from the time-evolution picture may not be enough to save traditional accounts of lawhood: the increasing importance of non-dynamical approaches in physics gives us good reason to take seriously the possibility that  the next generation of fundamental laws won't be parsed within the kinematics/dynamics framework at all, and thus we are in need of some alternative account of lawhood which does not lean so heavily on that distinction.

\subsection{Retrocausal approaches} 

Retrocausal approaches to the interpretation of quantum mechanics have been attracting significant attention in recent years; see for example the two-state vector interpretation\cite{Aharonov}, the transactional interpretation\cite{Cramer}, Kent's approach to Lorentzian beables\cite{Kent}, Wharton's retrocausal path integral approach\cite{Wharton_2018}, and Sutherland's causally symmetric Bohmian model\cite{Sutherlandretro}. The proliferation of such models provides good reason to suppose that a correct understanding of quantum mechanics may require us to rethink some of our ideas about time and temporal evolution. Of course, one can always choose to rewrite retrocausal laws in a non-retrocausal way by  adding appropriate hidden-variables and/or fine-turning initial states; but it has been argued that in certain cases the resulting non-retrocausal law will fail to obey time symmetry\cite{Price2012-PRIDTI-2, PuseyLeifer} or will contain inexplicable coincidences\cite{Evans_2013}  and thus even if the non-retrocausal description always remains technically possible, there are compelling reasons to take the retrocausal descriptions seriously. 

As noted in ref \cite{Adlamspooky}, there are two importantly different notions of retrocausality currently employed in the field. Some approaches (such as the two-state vector interpretation) explicitly postulate two distinct directions of dynamical causality, such as a forwards-evolving state and also a backwards-evolving state, and in such cases we might hope to accommodate retrocausality within standard pictures of lawhood by making a small alteration which allows for evolution backwards as well as forwards. But other approaches (such as Wharton's models) are more at home within the kind of `all-at-once,' block universe model that we have already encountered in section \ref{Lagrange}. Moreoever, there are strong indications that the latter approach is the more coherent one: ref \cite{Adlamspooky} argues that only the `all-at-once' picture can circumvent the conceptual difficulties associated with retrocausality, and ref \cite{doi:10.1063/1.4982766} argues that even in accounts which appear to promote two directions of dynamical causality, `\emph{such dynamical stories are just narrative overlays on a static ontology}.' Thus these developments in physics suggest strongly that we are in need of new philosophical approaches to lawhood which are better aligned with the `all-at-once' picture.

\subsection{Quantum information and foundations \label{QI}} 

Large sectors of the growing field of quantum information science are concerned with discovering general constraints on what information-processing tasks can be achieved using quantum systems. For example, there is the `no-signalling principle' - that is to say, the fact that measurements on distinct quantum systems commute entails that quantum mechanics can't be used to send signals faster than light\cite{MRC}. A generalisation of no-signalling known as `information causality' has been used to derive the exact value of the limits on quantum non-local correlations\cite{Pawlowski}. We also have `monogamy,' which places limits on how strongly quantum systems can be entangled with two or more distinct systems\cite{Toner, Toner2}, and `no-cloning,' which places limits on the ways in which quantum information can be copied\cite{Scaranicloning}. These sorts of constraints are regarded by many researchers in the field as being deep and fundamental features of physical reality, and yet they are certainly not dynamical laws or time evolution laws in the usual sense, since they are primarily concerned with describing what is possible or impossible.

Similarly, in the adjacent field of quantum foundations,  there has been a great deal of interest in	operational theories\cite{CDK, PhysRevA.81.062348} and a related family of research programmes including nonlocal correlations\cite{Rohrlich, Pawlowski, Toner}, generalized probabilistic theories\cite{Hardyreasonable, Barrett, SpekkensBarrettLeifer, Masanes_2011}, empirical models\cite{Abramsky_2011}, and device-independent approaches\cite{newlabelColbeck, AdlamKent2, kthw,Acin}. These frameworks are often used to prove relations of the form `every theory which has property $A$ also has property $B$,' or to come up with operational axiomatisations of quantum mechanics. The properties and axioms that feature in these sorts of approaches are, once again, not usually concerned  with dynamics or time evolution: they frequently take the form of constraints delimiting the the bounds of the possible, such as `the existence of a continuous reversible transformation between any two pure states'\cite{Hardyreasonable} or `the existence of an information unit.' \cite{Masanes_2011} Again, proponents of these approaches often seem to be of the opinion that these properties and axioms are deep, fundamental features of the physical world, but they are clearly not dynamical or time evolution laws. Thus it seems desirable to have some kind of framework in which we could at least entertain the possibility that these sorts of information-theoretic and operational constraints might be genuinely lawlike. 

\subsection{Consistency Conditions}

Einstein's equations of general relativity admit solutions which contain closed timelike curves, i.e. points in spacetime which lie in their own causal future\cite{Thorne:1992gv}. The existence of closed timelike curves raises the spectre of logical paradoxes like the grandfather paradox, where someone travels into the  past via a closed timelike curve and shoots their grandfather in order to prevent their own conception. To rule out the possibility of such paradoxes it is necessary to impose consistency constraints on the fields at points along the closed timelike curve. Earman  considers that the best way of conceptualizing these consistency constraints is to suppose that they  `\emph{have law status}' and moreover, he argues that it's unlikely these constraints can be expressed in wholly local terms, and thus \emph{`in general, the consistency constraints may have to refer to the global structure of spacetime.'}\cite{Earman1995-EARBCW} Therefore if consistency constraints are indeed laws, it's likely that they are not local time evolution laws but rather some new sort of global, all-at-once laws. Earman suggests understanding these sorts of  laws within the best-systems paradigm (see section \ref{Hume}, but in principle there is no reason why one could not entertain global laws of this kind within a realist approach to lawhood, and thus there is a clear need for a realist framework which accommodates constraints of this kind. 

\subsection{Constraints \label{Con} }

 The notion of expressing laws as constraints has attracted interest in other areas of physics as well. For example, Deutsch and Marletto's constructor theory expresses physical law in terms of what tasks are possible and impossible\cite{Deutsch_2015}. The second law of thermodynamics is a law of this kind - in the Kelvin formulation, it tells us that processes which convert heat completely into work are \emph{impossible} in the constructor theory sense, i.e. the conversion of heat into work might occasionally occur by fluke but it can't be reliably achieved. Note that the concepts of `possible' and `impossible' employed here are in a sense non-local, since they refer not to individual instances but to behaviour over repeated instances:   \emph{possible} tasks are those which can be executed repeatably, in a cycle, with arbitrary accuracy, whilst  \emph{impossible} tasks are those which can't, though they might still occur as a one-off. Moreover, we know from past failed attempts at deriving the second law from the time evolution equations used in statistical mechanics\cite{BROWN2009174,Valente2014-VALTAT-7} that it is typically very difficult to account for such impossibility constraints in terms of purely time-evolution laws, so it seems likely that in order to take Deutsch and Marletto's construction seriously we will need an account of lawhood which is not predicated on time evolution.

 Similarly, Filomeno discusses various phenomena within physics where the dynamics exhibit typicality properties such that we get emergent stable patterns which are almost independent of the details of the underlying dynamics, and therefore it is the \emph{constraints} on the kinematically possible states (e.g. symmetries of the state space) which really determine the behaviour. Filmeno thus argues that we can explain the existence of regularities by committing to modal constraints on the kinematic state space rather than being ontologically committed to specific dynamical laws\cite{Filomeno2021-FILTOD}.  Constraints of this sort are clearly not time-evolution laws, and although we might attempt to rewrite them in that form, doing so would seem to erase the specific modal content of the claim that some sort of process is either possible or impossible. Constraint-based laws seem much more at home within some sort of block universe picture, where we can imagine them applying all-at-once to prohibit the relevant sort of process across the whole of history. Thus, given the increasing importance of constraint-based language in physics, it seems reasonable to require that our account of lawhood should allow the possibility of laws which take this sort of form.

\section{Existing Accounts \label{EA}} 

The examples of section \ref{intro} reveal a general trend toward laws of physics which don't fit into a traditional time-evolution picture - laws which apply `all-at-once,' which are non-dynamical, which are phrased in the language of constraints. How do traditional accounts of lawhood fare with these sorts of laws?  In this section, we will examine several popular  views and it will become clear that none of them is sufficient to accommodate the wide variety of non-Newtonian laws described in section \ref{intro}.
 
 \subsection{Humeanism \label{Hume}} 
 
Humeanism is the name given to a family of metaphysical attitudes which insist that nothing exists other than the `Humean mosaic,' i.e. the set of local matters of particular fact instantiated within our world, and that therefore all modal features of reality simply supervene on the physical. The most common Humean approach to lawhood is the best-systems view, which  holds that laws of nature are simply the axioms of the best systematisation of the Humean mosaic, where the `best systematisation' is supposed to be the one with the best combination of simplicity, robustness, and strength\cite{Lewis1980-LEWASG}. The best-systems account thus tells us that laws are merely \emph{descriptive} and have no modal force of their own. 
 
 The best-systems approach does accommodate a fairly large range of possible laws, as it allows that anything which can be written as the axiom of a best-system is a potential law. But there is one sort of case where the Humean may struggle - laws dealing with possibility or impossibility, as discussed in sections \ref{QI}, \ref{Con}. It's difficult for the Humean to do justice to the modal force of such laws, given that a central tenet of the Humean approach is the denial of modal concepts: the best option within the best-systems picture appears to be analysing `X is possible' as `X occurs' and `X is impossible' as `X does not occur.'  But in some cases we might want to say that `X is possible' is a law even though X does not actually occur.  For example,  in several of the axiomatisations discussed in section \ref{QI} it is shown that the fundamental structure of quantum mechanics can be derived from a set of axioms including \emph{continuous reversibility:} `there exists a continuous reversible transformation between any two pure states,' (i.e. every transformation between pure states \emph{ is possible} in both directions )\cite{Hardyreasonable, Masanes_2011}. Yet it may well be the case that not every transformation between pure states is actually instantiated in our world - indeed quantum mechanics defines an infinite number of such transformations, so the actual Humean mosaic can instantiate them all only if it contains an infinite number of events, and we are not yet sure whether or not the actual Humean mosaic is infinite.  So it's likely that in this sort of case the analysis of `X is possible' as `X actually occurs' is not available, which makes it hard for the Humean to maintain that a principle like continuous reversibility could possibly be lawlike. Yet the axiomatisations suggested in ref \cite{Hardyreasonable, Masanes_2011} give us at least heuristic grounds to argue that the lawlike nature of this possibility statement is an important part of the reason why our world obeys quantum mechanics; it isn't my intention to claim here that this argument is necessarily correct, but arguments of this sort are taken seriously by at least some sectors of the physics community and therefore it's important that our account of laws of nature should be capable of making sense of them. Humeanism therefore seems a poor fit for at least some directions that the physics community is currently pursuing. 
 
 In any case the Humean approach is not a realist approach to lawhood - that is, it does not allow us to say that laws \emph{make things happen} and have modal force of their own. Yet the governance paradigm for lawhood still has a high degree of intuitive plausibility and moreover presents an importantly different picture of the relationship between laws and reality which may offer alternative insight into some of the philosophical topics which make use of the notion of lawhood. So even if it ultimately turns out that the Humeans can solve the problem with possibility-based laws and accommodate the full spectrum of laws postulated in modern physics, there would still be good reasons to carry on looking for a  \emph{realist} account of lawhood which can do the same, where by `realist' we mean an account which holds that modality does not simply supervene on the physical and laws form part of the objective modal structure of reality. Let us therefore proceed to realist approaches,

\subsection{Universals}

One of the most prominent realist approaches to the analysis of laws, developed by Armstrong\cite{Armstrong1983-ARMWIA}, Dretske\cite{10.2307/187350}, Tooley\cite{10.2307/40230714} and others, tells us that laws are relations between universals. In particular, Armstrong's account depends on a concept of universals as \emph{perfectly natural properties},  so a law is to be understood as a special necessitation relation that holds only between perfectly natural properties. For example, Newton's second law $F = ma$ is to be analysed as the claim that a relation of necessitation holds between the universals `force,' 'mass' and `acceleration,' which all look superficially like good candidates to be considered perfectly natural properties.

Various objections to the universals account have been raised, but from the perspective of the present article, the most important problem  is that it has an unavoidably reductionist flavour, because `perfectly natural properties,' are those which `slice nature at the joints,' and this slicing metaphor is often taken very literally: the perfectly natural properties are considered to be the  most basic properties of the smallest things in nature, whatever they might be. Moreoever, this necessitation approach is most naturally situated in a context where we have a clear distinction between kinematics and dynamics, because the universals themselves typically describe possible features of kinematical states and the relations between them define the dynamical laws determining how these features evolve over time. So the universals account seems to work well if we limit ourselves to  dynamical laws governing the local evolution of the most fundamental entities, but starts to look fairly weak if we consider any other types of law. For example,  it has been argued that the universals approach struggles to account for conservation laws\cite{Bird2007-BIRNML-2, 10.1093/bjps/43.3.371}, functional laws\cite{Vetter2012-VETDEA}, laws concerning irreversibly defined properties\cite{Hicks2017-HICDPI-2} and laws about the retention of dispositions\cite{tugby2017problem}. And these are just laws which exist already within mainstream modern physics - if we try applying the universals picture to some of the emerging ideas discussed in section \ref{intro} we will undoubtedly have even greater difficulties.

One could, of course, attempt to modify the universals account to be less reductionist in character. This would involve making a change in what we consider to be candidates for `perfectly natural properties' - rather than including only basic properties of very small things, we should allow this category to include properties of spatially distributed entities and processes, such as `the Lagrangian.' It seems, however, that such a move would carry us too far from what we would naturally understand as a `universal,' or `perfectly natural property.' In particular, one obvious desideratum for a perfectly natural property is that it should not exhibit multiple realizability: it should be exactly the same kind of thing wherever it occurs. But many of the laws described in section \ref{intro} concern properties which \emph{do} appear to exhibit multiple realizability - for example, the no-signalling principle. We could perhaps try to write no-signalling as a law of the form  `(signalling) necessitates (timelike or lightlike connectedness),' but are `signalling' and `timelike or lightlike connectedness,' really universals?  `Signalling' is a general property that applies to any type of process which allows the transmission of information, and there are many different ways to achieve signalling -  electromagnetic waves, the spoken word, carrier pigeon, and so on. It seems implausible to suppose that there's any one `perfectly natural property' that all these very different processes have in common - certainly, they do not seem to satisfy the desideratum of `exact similarity.' In a similar vein, we also see multiple realizability in the phenomenon of `universality'  in physics - the fact that strikingly similar behaviours can be exhibited by physical systems of very different types\cite{kadanoff1990scaling}. One might reasonably hope to subsume all these different processes under a single governing law, but regarding these very different systems as instantiations of the same universal would seem to be straining at the boundaries of what we could reasonably regard as a `perfectly natural property.' Thus it seems unlikely that we can relax our definition of `universal' enough to accommodate the variety of laws which appear in modern physics without losing hold of what makes the account in terms of universals compelling in the first place.

\subsection{Powers \label{capacities}} 

An alternative realist account of laws of nature, due in particular to Cartwright, regards them as simply statements about the powers/capacities/dispositions of entities\cite{Bird2005-BIRTDC, Ellis2001-ELLSE-2, Molnar2003-MOLPAS, Mumford2002-MUMLIN, Bird2007-BIRNML-2}. These powers/capacities/dispositions are regarded as being real, irreducible, and causally responsible for the way in which things turn out: when entities interact, their individual powers jointly determine what the outcome of the interaction will be.

The power account does seem a little more flexible than the universals approach: for example, if we are willing to to accept claims about what powers entities do \emph{not} have as part of the powers account (a controversial point in itself), we could imagine expressing the no-signalling principle in the form `no entity has the power to produce superluminal signals,' However, this approach to lawhood is  highly  local in character, for the entities which are the bearers of these powers are usually supposed to be the most fundamental entities, whatever they might be - for example, one might refer to the power of one fundamental particle to attract another. So powers are to be attached to specific objects in specific local regions, which doesn't work well for laws applying in a non-local way across large regions of spacetime, such as constraint-based laws or the optimization of a Lagrangian. Moreoever, the plausibility of regarding laws as powers is dependent on a clear distinction between kinematics and dynamics, because our idea of `powers' is an essentially anthropomorphic one: we think of powers as being exercised at particular moments and we take it that the exercise of powers produces effects in a temporally ordered sequence. Thus it has been noted\cite{Ioannidis2021-IOANLA} that the powers account has difficulty dealing with conservation laws and symmetry laws, and the same seems true for other sorts of non-dynamical laws. It's not obvious to see how one could describe the optimization of a Lagrangian as a power: to which entity do we ascribe the power, and at which moment is this power exercised?  

Again, one could attempt to modify this account to make it suitable for a non-dynamical  context. This would involve making a change in what we consider to be the candidates for bearers of nomic powers - rather than including only very small things, we should allow this category to include spatially distributed entities and processes. For example, refs \cite{ 10.2307/687764, Bird2007-BIRNML-2} suggest that the powers approach should understand conservation laws and symmetries  as resulting from the `dispositional	essence	of	the	property of	being	a	world.'  However,  as argued in \cite{Ioannidis2021-IOANLA}, once one starts ascribing laws to powers of the `whole world' it's unclear what need there could be to invoke powers of individual entities at all, so the central thesis of the powers approach starts to look superfluous. Moreover, once we start talking about powers of the world as a whole, the idea of powers being associated with \emph{individuals} which act \emph{at times} on \emph{other individuals} loses all meaning, and it becomes unclear in what sense these `powers' can be described as such at all.  Similarly, it's unlikely that any generalisation of the powers view which entirely put paid to the distinction between kinematics and dynamics would involve anything that would look sufficiently like a `power' to merit the term; so continuing to describe this picture in terms of powers only serves as obfuscation.

It's also worth nothing that powers on their own are not adequate to reconstruct most scientific laws. For after all, idiosyncratic individual causal powers are not much use to us as epistemic agents; it is necessary that each instance of a particular type of entity should have broadly the same powers, so that we can, for example, draw conclusions about the behaviour of electrons in the future based on our observations of electron behaviour in the past. Different accounts achieve this in different ways - for example,  ref \cite{Ioannidis2021-IOANLA} offers an explicitly dualist account postulating both fundamental powers and fundamental laws governing the ascription of the powers to appropriate entities. The dualist approach does somewhat increase the flexibility of the powers account, making it possible, for example, to accommodate conservation laws, and therefore one might think that such a dualist account would give us the resources we need to model non-dynamical laws. But although this approach does allow some level of global coordination, ultimately it still requires us suppose that these global laws take effect only via individual local interactions:   `properties are	‘told’	what	to	do	by	the	nomic	relation,' \cite{Ioannidis2021-IOANLA} where the properties and the things that they do are understood to be local and specific to individual entities. So this dualist approach ultimately looks like an attempt to `localize' the global laws by siting their effects in properties of specific entities rather than allowing that laws may govern holistically and all-at-once; if we want an account of lawhood which accommodates genuinely global, atemporal laws, the  halfway house offered by the dualist account should not satisfy us. 

\subsection{Primitivism \label{prim}}

So far we have focused on reductionist accounts of lawhood, but an alternative exists in the form of primitivism, i.e. the idea that lawhood cannot be analysed at all and must be considered as a primitive fact. This approach is advocated by Maudlin\cite{Maudlin2007-MAUTMW}, Carroll\cite{carroll_1994}, and Lange\cite{lange2000natural} among others. Now, one of the difficulties associated with primitivism about lawhood is that it is by nature stipulative, and  this means that if we are not careful, we are in danger of simply building into the definition of lawhood  whatever assumptions about the laws of nature our ancestors happen to have made about laws. For example, although Maudlin believes that laws are primitive and unanalysable, he asserts that the fundamental laws are all `fundamental laws of time evolution'\cite{Maudlin2007-MAUTMW} and thus the time evolution assumption is built into his concept of a primitive law, so Maudlin's particular take on anti-reductionism with respect to laws looks to be incompatible with many of the types of laws we considered in section \ref{intro}. Of course other anti-reductionist approaches would not necessarily have this feature, but they might build in other  assumptions: the general point is that if we are not careful, the assertion that laws are unanalysable will  just amount to an excuse for being dogmatic about what laws look like. 

That said, there is no reason why a sufficiently general primitivist account could not provide a good way of thinking about non-dynamical laws. Indeed, the approach we advocate here might be regarded as a version of primitivism.  We do not intend to be  fully committed to primitivism because we don't necessarily want to claim that no correct reductionist account of lawhood could possibly exist;  we are simply agnostic on this point, and moreover even if such an account did exist we are sceptical that we could ever find out what it is. However, the distinction between this position and primitivism is fairly small, and thus the framework that we set up here is likely to be of interest to primitivists too. A primitivist could certainly maintain that ultimately laws are primitive entities, but hold that the  framework introduced in this paper serves to `axiomatize the primitive' in Schaffer's sense\cite{boltg} by defining a general space of possible laws from which we may draw specific individual primitives. 

\subsection{Modal Structure \label{modal}}

Finally, let us consider the position that laws of nature should be regarded as part of the objective modal structure of reality. This approach to lawhood is most commonly seen in the context of structural realism, although it might also be regarded as a form of primitivism as discussed in section \ref{prim}. From the structural realist point of view there are two different ways to motivate this position, roughly corresponding to the two distinct strands of structural realism. First, one could argue that since we have access to laws only via their effects on the physical world, we have no epistemic access to them in and of themselves, and therefore we should be epistemically committed only to beliefs about the modal structures which laws induce and which mediate their effects on physical reality, rather than metaphysical claims about the underlying nature of the laws. This approach is similar to traditional arguments for epistemic structural realism (ESR)\cite{Morganti2004-MOROTP-9}. Alternatively, one could go a step further and argue that laws \emph{are} nothing above and beyond modal structure  and therefore by learning about those structures we have learned all there is to know about the laws. This approach is akin to ontic structural realism (OSR)\cite{Ladyman1998-LADWIS-2}. In this article we will make no attempt to adjudicate between these two positions: for present purposes the important point is that both approaches have the consequence that understanding laws comes down to understanding modal structures.

In order to see if this view can accommodate laws beyond the time evolution paradigm, it will be necessary to say something about what we mean by `modal structure.'  The most common approach in the literature conceptualizes laws in terms of \emph{causal} structure, and in accordance with the asymmetry of the causal relation, lawhood is then understood in a temporally directed way, such that laws `\emph{generate or produce}' later states from an initial state\cite{Maudlin2007-MAUTMW}. For example, ref \cite{Berenstain2012-BEROSR} describes modal structure  in terms of the existence of `genuine causal powers,' and ref \cite{Ladyman2007-LADETM} holds that these powers are typically supposed to flow `along asymmetrical gradients.' Similarly, ref \cite{newnamees} gives an account of modal structural realism in terms of causal structure, drawing on the well-established contrast between causal and categorical properties. But conceptualizing modal structure in this causal, dynamical, temporally directed way seems suited only for laws within the time-evolution picture, since many of the more general sorts of laws mentioned in section \ref{intro} have no built-in temporal or causal direction. Indeed, in light of  developments in modern physics it would seem foolhardy for proponents of the modal structure approach to tie their position too strongly to \emph{causal} structure: if the `all-at-once' view of lawhood is accepted then there probably can't be anything like causal structure at the most fundamental level, since events determined by all-at-once laws will typically depend on one another in a reciprocal fashion, so there will be no asymmetrical causal relations to be found.  But the absence of specifically \emph{causal} structure need not defeat the modal structure approach to lawhood, because of course there could still be more general modal structure in the all-at-once setting (e.g. relations like metaphysical necessitation or ontological dependence), and therefore it should be a priority for proponents of modal structure to develop an account of these more general possibilities. 

Another popular option is to refrain from saying anything at all about the  specific modal structures associated with the laws of nature. This approach is linked to `selective realism,'  a form of realism which aims to overcome the pessimistic meta-induction by making epistemic commitments only to carefully selected features of our theories which we believe are very likely to survive all future theory change\cite{psillos2005scientific}. Structural realism is sometimes regarded as a form of selective realism, where the feature that is supposed to survive theory change is simply `structure,' broadly construed\cite{papineau1996philosophy}. And clearly in order to make selective realism work for a modal approach to laws it will be necessary to refrain from saying anything too definite about the modal structures associated with laws, because greater specificity leads to a greater risk that the  structures we have identified will not in fact survive theory change. Indeed, in order to rule out underdetermination, the selective realist about modal structure would have to insist that more or less all the modal structures that one could imagine associating with a given theory are in  fact  equivalent, which prevents them from actually specifying any concrete structures which might be involved in this form of structural realism. This leads to the kind of objection that Esfeld makes in ref \cite{pittphilsci9091}, where he argues that OSR is only a `partial realism' because it provides `only a general scheme for an ontology of the physical domain,' and hence it is too vague to say anything interesting about, for example, how the EPR correlations come about.

But as argued in ref \cite{Ladymanstruc}, structural realism does not \emph{have} to be a form of selective realism and nor does structural realism require the selective approach in order to overcome the pessimistic meta-induction:  it is possible to maintain that the success of past theories is explained by continuity of structure through theory change without necessarily being committed to the view that we can always identify \emph{which} specific structures  will survive \emph{future} theory change. And in fact, on top of the unsatisfying vagueness of the selective approach, there are a few specific reasons why selective realism would be the wrong route for the current project. First, moving away from time evolution is a significant conceptual shift, and an important part of our motivation for this project was to arrive at an approach to lawhood which allows us to be explicit about that shift: the selective approach to modality is too vague for this purpose. Second, another part of the motivation for this project was to make it possible to give generalised definitions of notions like determinism outside the time evolution paradigm; and if we are to do that by appeal to modal structure, we will require some fairly definite framework in which to characterise modal structure in order to describe features of different possible modal structures. Thus for our purposes the selective approach is not adequate: the ongoing research programme into laws outside the temporal evolution paradigm needs  a  more robust approach to modal structure. 

\section{Modal Structure} 

None of the realist approaches we examined in section \ref{EA} are at present a good fit for the kind of `all-at-once,' non-dynamical laws that we explored in section \ref{intro}. Thus our aim in this section is to develop a realist account which naturally accommodates these sorts of laws. In particular, we will argue that such an account can be achieved by employing the modal structure approach as described in section \ref{modal} together with a generalisation of the notion of modal structure to go beyond merely causal structure. 

We emphasize that this article is focused on articulating modal structural realism regarding specifically \emph{laws} - this approach could certainly be coupled with a  structural realism regarding objects in general, but it could also be paired with a more conventional object-oriented realism. For example, one could perfectly well hold that laws are nothing but modal structure whilst also maintaining that the things governed by these modal structures are ordinary physical objects, not structures. So although the position presented here may seem particularly at home within a structural realist setting, other sorts of realists are also free to adopt it. 
 
 \subsection{Constraints} 
 
  Taking inspiration from the examples in section \ref{intro}, we propose characterising the modal structure associated with laws in terms of the \emph{constraints} that the laws induce. Constraints are a suitable tool because the idea of a constraint (at least in the popular imagination) is a modal one: we typically think about constraints not merely as \emph{descriptions} of the way things happens, but as dictating the way things \emph{must} happen.  Indeed, Ross, Ladyman and Spurrett use language of this sort in ref \cite{Ross2007-ROSCIA-3}: `\emph{If there are structural facts about the whole universe, and these facts constrain all the facts about particular regions of the universe ... then the only necessity in nature is furnished by these constraints. The constraints - that is, the structures themselves - are real patterns.}' 

A constraint places limits on the way in which things are allowed to happen - or equivalently, it delimits those things which are \emph{not} allowed to happen. Constraints are therefore in some ways similar to powers -  the powers approach proceeds by enumerating all of those things which are possible  for certain sorts of systems, while imposing constraints similarly amounts to dividing a space of configurations or histories up into those which are possible and those which are not possible. However, thinking in terms of powers emphasizes the \emph{possibility} side of the division, whereas thinking in terms of constraints emphasizes the \emph{impossibility} side of the division, thus allowing us to take advantage of the simple logical point that if something is not possible it definitely will not occur, whereas if something is possible it may occur but it may also not occur. Laws defined in terms of what is \emph{not} possible thus have the advantage of being completely transparent in their empirical content: it is very straightforward to identify empirically whether or not a law of the form `X is impossible' is obeyed in any given situation, whereas there is no easy way to say whether or not a law `X is possible' has been obeyed, since the nonoccurrence of X does not entail the impossibility of X. This feature will allow us to define  the modal structure induced by constraints in a clear and empirically meaningful way. Moreover constraints, unlike powers, are not typically associated with individual physical systems: Lagrangian constraints, no-signalling constraints,  thermodynamical constraints and their ilk are applied to whole histories and to composite systems rather attached to objects and exercised at times, and therefore constraints are a suitable way of implementing the vision of all-at-once, non-dynamical laws presented in previous section. 

One might perhaps have the worry that laws which do nothing other than specify that certain things are not allowed to happen would not be sufficiently powerful to explain the wide range of lawlike behaviour exhibited in our world - surely we need a specification of what \emph{does} happen in addition to a specification of what cannot happen? But in fact, specifying what doesn't happen can also be used as a way of determining what will in fact happen.  A standard deterministic time evolution law which says that a certain sort of system always behaves in a certain way can be rewritten as a constraint which says that it is not possible for systems of that sort to behave in any other way - for example, the law $F = ma$ can be written as a constraint ruling out all accelerations other than $F/m$. Similarly, a law which says that a certain sort of system behaves in a certain way with high probability can be written as a constraint ruling out all histories in which the relevant behaviour occurs too infrequently. We will look at more examples in section \ref{example}, but for now it is enough to observe that specifying what is impossible will suffice to determine what does in fact happen provided that a large enough provided that a large enough range of possibilities are ruled out.
 
In accordance with the structural approach discussed in section \ref{modal}, we will refrain from speculating on the metaphysical nature of whatever it is that grounds constraints (if indeed anything does), and instead characterise them entirely in terms of the role they play in defining the objective modal structure of reality. First, clearly constraints are to be understood as distinct from the Humean mosaic (since they are ex hypothesi modal) but as playing some role in determining the contents of the Humean mosaic (since this is the function for which we have postulated them). That is to say, constraints are  \emph{ontologically prior} to the Humean mosaic. \footnote{To define ontological priority, we first define ontological dependence by saying that $A$ ontologically depends on $B$ if $A$ depends for its existence on $B$: for example, sets depend for their existence on their members. In many cases (indeed, some say \emph{all} cases) ontological dependence seems to be asymmetric - $A$ ontologically depends on $B$ but not vice versa - and in such cases we say that $B$ is \emph{ontologically prior} to $A$\cite{Schaffer2010-SCHTIR-6, 10.2307/40468234}. To enliven this rather abstract language,  ontological priority may be described metaphorically as specifying the order in which God would have to create things to make a world like ours\cite{Schaffer2009-SCHOWG}.} In order to make this analysis more concrete, we will henceforth define constraints \emph{extensionally}, appealing to techniques employed in modal logic: a constraint will simply be defined as a set of Humean mosaics, i.e. the set of all mosaics in which that constraint is satisfied. 

The structural role of a constraint is simply to require that the actual Humean mosaic must belong to the corresponding set, and in that sense a constraint is indeed ontologically prior to the actual Humean mosaic, since the constraint dictates the possible contents of the Humean mosaic.  Indeed, the extensional definition of constraints allows us to say something more concrete here. A constraint guarantees that the actual Humean mosaic must belong to the corresponding set; that is to say, every possible world in which a given constraint holds is a possible world in which the Humean mosaic of that world belongs to the associated set. So for example, if every mosaic in the associated set exhibits a certain sort of regularity, every possible world in which that constraint holds is a possible world in which that regularity is exhibited, which is the definition of metaphysical necessitation (in the possible world semantics, we say  that `$X$ metaphysically necessitates $Y$' if and only if `every possible world in which $X$ is true is also a world in which $Y$ is true.'). So constraints determine the contents of the Humean mosaic in virtue of \emph{metaphysically necessitating} that the actual Humean mosaic should have certain features. Metaphysical necessitation is clearly a modal relation, but there is no need for it be attached to individual local systems or to have any particular temporal direction, and therefore it is a good candidate for the kind of general modal structure that we need to express a large variety of laws outside the time evolution paradigm. 

This way of defining constraints is analogous to the idea that propositions can be defined as sets of possible worlds. Indeed, for the proponent of Humean supervenience who believes that a `possible world' just is a Humean mosaic,  the definitions are equivalent. However, the case of propositions is complicated by the fact that propositions have intensions, and it has been suggested that the definition in terms of possible worlds is too coarse-grained to respect this intensionality\cite{Stanley2010-STAAAI-7}. But since we intend to understand `constraints' as a matter of objective fact, we need not worry about intensions, so the view of propositions as sets of possible worlds fits our purposes very nicely: it is precise, clear, and allows us to set out a space of candidate laws which assumes very little about what those laws might look like. In some cases, a constraint will be expressible in simple English as a requirement like `no process can send information faster than light,' such that the  constraint corresponds to exactly the set of mosaics in which the requirement is satisfied. But there are many sets of mosaics which will not have any straightforward English characterisation, so we will not be able to state the corresponding constraint in simple terms. Nonetheless, each set still defines a unique constraint.

\subsection{Laws}

What then is the relationship between laws and constraints? Well, in many cases we can simply think of laws as inducing constraints - for example, if the no-signalling principle is considered to be a law of nature, it induces a constraint consisting exactly of all the non-signalling mosaics.  Such a law can be interpreted as saying that the actual Humean mosaic must belong to the corresponding constraint; and in the case where the constraint can be given a simple English characterisation, the law can be interpreted as saying that the actual Humean mosaic must have the corresponding property. If all of the laws of nature  induce constraints like this, then clearly the actual mosaic must lie in the intersection of all the constraints associated with the laws.

However, simply equating laws with constraints would rule out the possibility of laws which are genuinely probabilistic even from the external, all-at-once point of view, such as a law assigning a probability distribution over possible initial states of the universe. So for full generality, we will instead suggest that laws should be thought of as inducing \emph{probability distributions} over constraints. In simple cases like the no-signalling principle, where the law is  non-probabilistic, the corresponding  distribution will be trivial - i.e. it will assign probability one to a single constraint and zero to all other constraints. In other cases, the distribution associated with a law may be non-trivial, so we must imagine that first a constraint is drawn according to this distribution, and then the actual mosaic is required to belong to the corresponding set of mosaics. In this picture, we can think of a constraint being drawn for each law according to the probability distribution that the law induces, and then the actual mosaic must lie in the intersection of all of the selected constraints.

The idea of a constraint being selected according to some distribution may seem conceptually problematic, given that this process presumably only happens once and therefore no relative frequency interpretation is possible. However, sceptics about objective chances or single-case probabilities can always choose to insist that all the actual laws induce trivial probability distributions -  for as we will see in section \ref{probability}, it is perfectly possible to observe apparently non-trivial probabilities from our local point of view within a world governed by laws of nature which induce trivial probability distributions from the external point of view, so this option is not inconsistent with any current empirical evidence. Therefore scepticism about objective chances need not derail this approach: it merely places certain limits on the space of possible laws which we can contemplate.

Note that we could have chosen to work directly in terms of probability distributions over Humean mosaics, rather than constraints.  However, there are several reasons for preferring the constraint-based formulation. First, it is inspired by the increasing importance of constraint-based language in physics, as seen in section \ref{intro}. Second, it allows us to acknowledge that a law may require (deterministically or probabilistically) that the actual Humean mosaic belongs to a given set, but then say nothing further about which particular mosaic within the set will be selected - that is, the law prescribes no distribution within the set, not even the uniform distribution. For example,  we might consider analysing the no-signalling principle as a law which requires that  the actual Humean mosaic must belong to the set of non-signalling mosaics, but it does not  necessarily seem right to say that the law  assigns a uniform distribution over all non-signalling mosaics: in fact the law simply singles out the allowed set, it does not  tell us how to assign probabilities within that set. Of course we  may  still end up assigning some distributions inside the set of allowed mosaics: for example, we may start with some priors which lead us to assign  credences over Humean mosaics, and then when we learn the existence of some law which requires that the actual Humean mosaic belongs to some set, we eliminate all mosaics outside that set and renormalize our credences, leading to a distribution which is non-zero only inside the specified set. But the initial credences are nothing to do with  the law -  it is only the updating process which is dictated by the law, and hence the resulting distribution within the set of mosaics is a subjective probability  distribution based on our credences rather than  an objective chance distribution specified by a law. As we will see in section \ref{probability}, this distinction is important for the analysis of determinism and objective chance.

Since constraints need not be expressible in simple English, the probability distributions induced by various possible laws need have no simple English expression. Thus just as the best-systems picture allows anything that can be written as an axiom to be a possible law, defining constraints as sets of mosaics similarly allows us to define a  general space of law-candidates with very few restrictions on what sort of laws are allowed. It also easily accommodates laws with non-local effects, because a generic constraint created by selecting a set of Humean mosaics at random will induce massively non-local relationships between events: for example, if we know that the world is subject to a constraint consisting of exactly two mosaics which differ on matters of fact at exactly two spacetime points $(x, t)$ and $(x + \Delta, t + \Delta')$, then when we observe what actually happens at spacetime point $(x, t)$ we immediately know what happens at  $(x + \Delta, t + \Delta')$ even if it is very far removed in space and time.

\subsection{Governance}

As noted above, in this setting the relation of `governance' is simply equivalent to metaphysical necessitation, and this provides us with an interesting new approach to the question of what it is for a law to govern. Within traditional time evolution pictures of lawhood we must envision a set of laws external to physical reality which somehow `reach in' and at each moment of time push particles along certain trajectories, or induce states to evolve in certain ways, or similar; whereas in the constraint-based picture  the laws of nature select a set of constraints, which single out a set of `allowed' mosaics, and then the actual Humean mosaic is selected from this set. So rather than reaching into the universe to meddle in the course of physical events, the modal structure associated with laws can be regarded as simply selecting Humean mosaics all at once in an external fashion, making lawhood consistent with a `block universe' picture since the whole mosaic is now the object of the law. Thus as hoped, this approach thus represents a concrete way of realising the `all-at-once' vision postulated in refs \cite{Wharton_2018, Wharton, Adlamspooky, QMG}.

Note that we have not attempted to provide any account of the mechanism by which the laws single out some set of mosaics, nor how the actual mosaic is drawn from this set. One could of course come up with a literal reading of this framework akin to Lewis's literal view of possible worlds\cite{Lewis1986-LEWOTP-3}, where we take it that the other possible Humean mosaics actually exist and so, in the deterministic case where laws are just constraints, laws are literally sets of  mosaics which  actually exist. However this would seem to undermine the idea that laws are ontologically prior to the actual Humean mosaic, as  the actual Humean mosaic is a part of all the sets which contain it, and thus we could only say that these sets are ontologically prior to the actual Humean mosaic if we were willing to say that sets are ontologically prior to their members, which would certainly be contentious\cite{Schaffer2009-SCHOWG,}. And in any case, a major motivation for moving to the structuralist picture is to avoid making specific commitments to the metaphysical nature of whatever it is that induces the modal structure in question (if indeed anything does). 

Nonetheless, even if the selection of the actual mosaic  from the candidate set is regarded only as a formal device, it still points to a different way of thinking about how laws govern, and therefore it offers an interesting opportunity to revisit some discussions around the notion of governance (e.g. see refs \cite{Ioannidis2021-IOANLA,Schaffer2016-SCHIIT-11, Hildebrand2019-HILSPA-3}). For example, traditionally proponents of OSR have been inclined to resist the notion that laws `govern,' because the idea of postulating a governing force independent from nature seems incompatible with the naturalism which motivates OSR. The constraint approach presented here addresses this concern by providing us with a naturalised notion of governance, where laws `govern' in exactly the same way that ordinary modal properties `govern' - a cause can be said to `govern' its effect, for example. Since proponents of OSR are typically willing to countenance modal notions and modal structure, this is a notion of governance which should be acceptable to them; although the constraints are ex hypothesi external to the Humean mosaic, they are not outside of or independent from nature as a whole, because according to the OSR worldview the objective modal structure of reality is an intrinsic part of nature. Thus this way of thinking makes it possible to effect a reconciliation between the traditional `governance' picture and the naturalised modal structure picture promoted within OSR.

Second, this approach might take the edge off some worries about how it is that laws operate in a consistent way across time. For if we take it that the laws of nature apply all at once to the whole of history, there's some principled reason why we should expect to see consistency across the whole of history; conversely, if laws operate locally and instantaneously as in the standard kinematical/dynamical picture, it's less clear why we should expect the laws to work in exactly the same way at every separate instance, and thus the worry that they might suddenly stop working somewhere in the future has more bite. Of course this is not intended to be a deductively valid argument: the constraint framework can't \emph{guarantee} that the future will be the same as the past, because laws which change discontinuously at a certain time or which fluctuate across history can certainly be expressed as constraints, and thus we still have to make some sort of consistency assumption. But at least in the global setting the laws are applied externally once and for all when the actual Humean mosaic is selected, so we don't have to worry about the possibility that the laws will suddenly stop working  or wonder about what coordinates different instances of nomic powers across spacetime.

\subsection{Humean Supervenience}

Because this framework is explicitly modal in character, it is of course not compatible with Humean supervenience: even though laws are characterised in terms of their effects on the Humean mosaic, it is nonetheless the case that two worlds with different laws of nature could give rise to the same Humean mosaic. This is true even if we limit ourselves to non-probabilistic cases where the mosaic is singled out uniquely by the laws of nature. For example, if $A$, $B$ and $C$ are Humean mosaics, one world could have a law of nature assigning probability $1$ to the single constraint $\{ A\}$ and a second world could have two laws of nature with the first assigning probability $1$ to the constraint $\{ A , B\}$ and the second assigning probability $1$ to the constraint $\{ A, C\}$, and the different laws of nature of both of these worlds will deterministically produce the same Human mosaic, $A$. So according to this realist approach, two worlds with the same Humean mosaic may differ in virtue of being governed by different laws. Indeed, there are an infinite number of constraints which would uniquely single out the mosaic $A$, since there are an infinite number of Humean mosaics from which we could select mosaics $B$, $C$ in order to form the constraint2 $\{ A , B\}$,  $\{ A, C\}$. But according to the approach we have set out here, there is exactly one set of constraints which does \emph{in fact} result from the actual laws of nature, and it is this unique set of constraints which represents the objective modal structure of reality, even though it is not the only set of constraints which is compatible with the  actual Humean mosaic. This is of course to be expected from any non-Humean approach to modality: if we reject Humean supervenience then it follows that modal structure can't simply be read off the Humean mosaic, and therefore it must be the case that not all of  the  possible modal structures which are compatible with the actual Humean mosaic are correct representations of the  objective modal structure of reality.  

\subsection{Humean Mosaics \label{mosaic}} 

So far our discussion has left the notion of a `Humean mosaic' largely unanalysed. But it is important to be clear that our use of Humean mosaics is not supposed to be a statement about our ontological commitments: although the Humean mosaic consists entirely of instantiations of properties across spacetime, we don't mean to claim that this minimal Humean ontology necessarily includes everything that exists. The real fundamental ontology could be objects, fields, processes, relations, structures, or indeed there could be no fundamental ontology at all: this approach is intended to be agnostic as to that point. But any fundamental ontology must ultimately give rise to some distribution of categorical properties across spacetime (even if `properties' and/or `spacetime' are regarded as emergent features), and this distribution of properties across spacetime is the level at which we as observers make contact with objective reality, so it is the appropriate level at which to describe the modal features of laws. Any two laws which gave rise to different modal structure for the fundamental ontology but identical modal structure at the level of the Humean mosaic would not be interestingly different from the point of view of human observers and our epistemic access to reality. 

 For example, proponents of OSR presumably would not wish to say that nature contains categorical, non-relational properties at a fundamental level, but they will nonetheless need to provide an account of how structures give rise to (at least the appearance of) distributions of properties across spacetime, and thus they can still adopt the constraint framework at the level of these emergent distributions of properties. Similarly, there are indications from quantum gravity that `spacetime' may be emergent from some other sort of substratum\cite{Rovelli2008}, and therefore we might not want to be committed to the claim that the spacetime on which the Humean mosaic is defined is genuinely fundamental. But laws governing this substratum will  have consequences for the geometry of spacetime and the distribution of properties across it, and therefore these laws will still give rise to meaningful constraints within the constraint framework. (Of course, given that it has so far proven difficult to use theories of quantum gravity to make many concrete empirical predictions, it might not be straightforward to determine what these constraints look like - but at least in principle they must exist!).

Even with this proviso, the sorts of laws which one can imagine coming up with will depend sensitively on what sorts of categorical properties one allows that the Humean mosaic will contain. For example, some physicists would probably insist that the Humean mosaic should contain properties like `spin' whereas others might feel that spin is an abstraction which can be completely accounted for by keeping track of spatial position (as is done in the de Broglie-Bohm interpretation\cite{tastevin2021outcomes}). This is in fact an intentional feature, because the project we are engaged on is motivated in large part by the naturalistic idea that a philosophical account of laws should be led by science and not vice versa, and as part of this vision it's important to leave open the question of which categorical properties appear in the Humean mosaic; this allows us to postulate a space of Humean mosaics which contain all of the basic properties featured in the theories we currently take seriously, whilst also allowing the possibility of mosaics which contain properties that we have not yet imagined. Indeed, specifying what sorts of categorical properties may appear in a Humean mosaic is loosely analogous to specifying an ontology, or to setting out the kinematical space of the theory, with further constraints that single out a smaller subset of mosaics then playing the role of the dynamics. Moreover, this specification can itself be regarded as a constraint: limiting ourselves to consideration of Humean mosaics which contain only pointlike masses is functionally equivalent to imposing a constraint consisting of all and only those Humean mosaics which contain only pointlike masses. So in principle we can imagine that the infinite space of Humean mosaics contains every possible mosaic featuring every possible categorical property, and the `laws of nature' then specify what sorts of properties exist as well as how those properties behave.

One type of realist view which might seem a little more difficult to accommodate in the constraint framework is a branching-world approach like the Everett interpretation of quantum mechanics\cite{Wallace}, since a single Humean mosaic doesn't seem adequate to fully capture what the world is like in such a setting. However, the role of the Humean mosaic in the constraint framework is simply to capture  the `actual, non-modal' contents of reality, in order that we can identify sets consisting of different ways the non-modal contents of reality could be. And presumably there is at least in principle some way of specifying the actual, non-modal contents of reality within an Everettian universe - for example, one might imagine replacing a `Humean mosaic' with a `universal wavefunction' and then consider a constraint to be a set of different `universal wavefunctions' with differing local or global structure. So the constraint framework could in principle be expanded to accommodate branching-world pictures. However, we have refrained from doing so here because the way in which human observers actually make contact with and learn about the objective modal structure of reality is via observations of what appears from our point of view to be a single Humean mosaic, i.e. various instantiations of properties across spacetime. Even if we are in an Everettian universe we will never be able to directly observe the structure of the universal wavefunction; at best we can form hypotheses about it based on what prima facie appears to be a single Humean mosaic. So the attempt to make inferences about the objective modal structure of reality encounters additional epistemic difficulties in the context of branching-world pictures, and we have elected here not to engage in detail with these issues - though we would invite proponents of branching-world or multiple-world pictures to take up these questions. 

\section{Examples \label{example}}

 In this section, we give some examples of how various different sorts of laws would look within the constraint framework.

\subsection{Time Evolution Laws} 

Our stated aim for this analysis was to provide a realist approach which is \emph{more} general than theories based only on the time evolution picture, so it is important that our framework should also  accommodate time evolution laws as a special case. It might seem a little odd to think about a time evolution law in terms of constraints on an entire Humean mosaic, since we typically think of time evolution laws as acting moment-by-moment on instantaneous states -  but in fact, many sectors of the philosophy of physics community have already adopted a constraint-based approach to time-evolution laws. For example, in ref \cite{pittphilsci16622}, Wallace writes 'A \emph{history} of the system is then a smooth function $q : \mathbb{R} \rightarrow Q$ assigning to each time $t$ the configuration $q(t)$ of the system at that time. The dynamical equations of Newtonian mechanics distinguish dynamically possible from dynamically impossible histories.' Similar language also appears in refs \cite{pittphilsci17184, List_2019}. In this sense, the proposal we have made here simply formalises a way of thinking about lawhood that has been common for some time but which has not yet passed over into mainstream philosophical analyses of lawhood (or at least, not into non-Humean analyses of lawhood). 

This approach allows us to rewrite any time-evolution theory in the form of a constraint theory where the role of the laws is not to act moment-by-moment on states but rather to pick out the set of possible histories. It is then straightforward to write such laws in the constraint framework: a deterministic evolution law which applies to systems of type $s$ is analysed as a probability distribution which assigns probability $1$ to the constraint consisting of the set of all Humean mosaics in which all systems of type $s$ have histories which are dynamically possible according to the law (where the `histories' in question may extend over the whole of time, or the entire time during which the system in question exists). 

We can also write indeterministic time evolution laws in a similar  language. In fact, there are two possible routes for probabilistic laws. The first option involves constructing a probability distribution over each possible history of a given system by multiplying the probabilities assigned by the law at each time-step, and if necessary taking a limit as the time-steps approach size zero. Then a law which assigns probability $p_s(h)$ to history $h$ for a particular system $s$ can be analysed as a law which assigns probability $p_s(h)$ to the set of all mosaics in which system $s$ has history $h$, and so on for every system of the same type across the whole Humean mosaic. Alternatively, we can implement a form of nomic frequentism\cite{Roberts2009-ROBLAF} where a law of the form `process $P$ produces outcome $A$ with probability 0.8 and outcome $B$ with probability 0.2,' is rewritten as the constraint  `across all occurrences of process $P$ over all of spacetime, 80 percent have outcome $A$ and 20 percent have outcome $B$.' Provided that the number of instances of process $P$ across spacetime is sufficiently large, the effect of this sort of constraint will be locally indistinguishable from the effects of constraints assigning probability distributions over histories. We will discuss these options further in section \ref{probability} but for now it is enough to observe that as previewed earlier,  both deterministic and probabilistic time evolution laws can always be rewritten in terms of constraints which rule out behaviour incompatible with the law. 

\subsection{Lagrangians}

A deterministic Lagrangian theory can be  written in the constraint framework  as a law which assigns probability $1$ to the set of all mosaics in which systems always take the path which optimizes their Lagrangian; and a probabilistic Lagrangian theory which assigns probability $p_s(h)$ to history $h$ for a particular system $s$ can be analysed as a law which assigns probability $p_s(h)$ to the set of all mosaics in which system $s$ has history $h$. This description allows us to take seriously the atemporal character of these descriptions: rather than saying that earlier parts in the history cause later parts and not vice versa, as in the traditional time-evolution picture, we are now able to allow that the laws select the whole history all at once, so the relationship between earlier parts and later parts of the history is reciprocal and holistic. 

\subsection{Retrocausal Laws}

We noted earlier that it is important to distinguish between the type of retrocausality which involves both forwards and backwards evolution, and the type of retrocausality which arises naturally in an `all-at-once' picture without either forwards or backwards evolution. The former can be accommodated within the constraint picture in the form of two distinct laws. In the deterministic case, for a retrocausal description of a system of type $s$, we have one law which induces a distribution assigning probability $1$ to the constraint consisting of the set of all Humean mosaics in which all systems of type $s$ have histories which are dynamically possible according to the forwards-dynamics, and another which induces a distribution assigning probability $1$ to the constraint consisting of the set of all Humean mosaics in which all systems of type $s$ have histories which are dynamically possible according to the backwards-dynamics. The probabilistic case is similar, except the laws would now induce nontrivial probability distributions assigning probabilities to histories according to the probabilities ascribed by the forward and backwards dynamics respectively. 

For the latter, we need only a single law which singles out a set of Humean mosaics exhibiting appropriate correlations between events at different times. As a simple example, consider a law inducing a distribution assigning probability $1$ to the set of all Humean mosaics in which the result of the measurement at position $(x, t)$ is the same as the result of the measurement at position $(y, t + \delta)$ with $\delta > 0$. If we happen to know about this law, then when we observe the result at $(x, t)$ we know immediately the value of the result at $(y, t + \delta)$, but it is not the case that the earlier result causes the latter and not vice versa: the law simply requires that they should be the same, and thus from the external, all-at-once point of view their effect on one another is perfectly reciprocal. Obviously in this very simple case there are no compelling theoretical reasons to prefer the all-at-once approach to the approach where the earlier result causes the latter one to have some value, but as noted in section \ref{intro}, in more complex cases we there are indeed several arguments in favour of the all-at-once approach, and the constraint framework allows us to postulate such all-at-once laws in a straighforward manner. 

\subsection{Information-Theoretic Constraints}

Clearly,  laws which are already written explicitly in terms of constraints, such as  the information-theoretic constraints discussed in section \ref{QI}, will be a natural fit for this framework. Moreover, if the law is already written in explicitly operational terms it is straightforward to identify sets of Humean mosaics obeying the law, since operational language can straightforwardly be translated to assertions about distributions of categorical properties over spacetime. For example, provided we can give a suitably operational definition of terms like `signal' and `faster than light,' then the `no-signalling principle' can straightforwardly be analysed as a law which assigns probability $1$ to the set of Humean mosaics in which no signal travels faster than light.

\subsection{Consistency Conditions} 

The case of consistency conditions for spacetimes without closed causal loops is a particualrly interesting one for the constraint framework, because consistency conditions don't actually have to be imposed as laws at all within the constraint framework: they emerge as consequences of the framework itself. This is because the laws of nature are understood to take the form of sets of Humean mosaics, or probability distributions over each set; and of course, each Humean mosaic represents a complete and consistent way the arrangements of physical fact within a world could be. Thus it is never possible to obtain a consistency problem from a law of nature which takes the form `the actual Humean mosaic must belong to a given set,' since inevitably every mosaic in the set is consistent. Consistency conditions are therefore a natural feature of this framework rather than a distinct set of laws which must be imposed externally. 

Moreover, it should be clear that this is the way things must be for any account of lawhood which does not uphold Humean supervenience. This can be seen from the following argument, where we postulate that laws and consistency conditions are independent and then derive a contradiction:

\begin{enumerate} 
	
	\item There exists a world $W$  governed by a set of laws $\{ L \}$ plus a law $C$ enforcing a consistency condition
	
	\item The law $C$ is logically independent from the laws $\{ L\}$
	
	\item If law $C$ did not hold in world $W$, the laws $\{ L\}$ would give rise to inconsistencies
	
	\item The laws of a given world do not supervene on the Humean mosaic of that world 
	
	\item 1), 2) and 4) together entail: there is another possible world $W'$ which is the governed only by the set of laws $\{ L\}$ without the consistency condition $C$
	
	\item 3) and 5) together entail: there exists a possible world $W'$ which contains inconsistencies
	
	\end{enumerate} 
	
But an `inconsistency' here refers to a scenario in which one and the same quantity has two different values at the same spacetime point, which is logically impossible (unless we are willing to countenance branching world scenarios). So, if we are not willing to countenance branching world scenarios, there cannot be a possible world which contains inconsistencies; thus we have derived a contradiction. It follows that any realist about laws (i.e. anyone who does not believe in Humean supervenience with respect to lawhood) must maintain that if a consistency condition is really indispensable, that consistency condition can't be independent from the laws themselves. The constraint framework delivers this guarantee, since no set of laws expressed within the constraint framework can ever lead to inconsistencies.

\subsection{Constructors}

Constructor laws are a little more complex than some other sorts of constraints, since they allow that `impossible' processes may occur as a one-off, so we can't simply analyse a constructor law as a law assigning probability $0$ to the set of all mosaics where the `impossible' process occurs. Rather, we must analyse such a law as a probability distribution assigning probability $0$ to the set of all mosaics in which process $x$  occurs \emph{in a cycle} - i.e. where the same setup is used to execute process $x$ some nontrivial number of consecutive times (and each execution is successful, or highly accurate). Fortunately this is possible within the constraint framework because of the non-local character of constraints, which allows us to consider laws as applying `all-at-once' to some extended series of events, rather than simply producing the events one at a time in a temporally ordered sequence. We would simply have to specify how often the process is allowed to occur at a given level of accuracy before we would consider the law to be violated, and then define a constraint consisting of all mosaics in which the number of occurrences of the process is smaller than this cutoff.

\section{Applications \label{applications}} 

As noted earlier, one of the main purposes of this framework is to facilitate re-assessements of the role played by laws in various areas of science and philosophy. In this section we briefly preview a few possible directions for future research. In particular, we reinforce the ways in which the realist approach set out in this paper leads to analyses that differ from comparable Humean accounts. It is not my intention to argue here that the realist analyses are necessarily superior to the Humean ones, but they are certainly \emph{different} analyses, and it is therefore useful to have access to this realist framework at least for the purpose of comparing and contrasting the two approaches.

\subsection{Realism vs Humeanism}

For many people it seems very  natural to suppose that laws \emph{govern} - that is to say, laws make things happen and are thus part of the objective modal structure of reality.  However, this realist picture of lawhood has to a large extent fallen out of favour in philosophy of science, and a significant proportion of modern philosophers favour the best-systems view. Why do philosophers find the best-systems account so compelling? There are of course a number of epistemic and metaphysical reasons which might lead one to question the existence of objective modal structure, but on top of these considerations the best-systems view also has one very important advantage: it tells us that anything  which could be written as an axiom of some `systematisation' could in principle be a law, and therefore it does reasonably well at accommodating  a variety of nonstandard laws that appear in modern physics (though as we noted in section \ref{Hume}, there may be difficulties with laws pertaining to possibility and impossibility). Conversely we saw in section \ref{EA} that existing realist accounts aren't typically very friendly to laws outside the time evolution paradigm, and thus any philosopher who accepts that an account of laws of nature should be capable of accommodating the full range of laws entertained by modern physics seems to have good reason to prefer the Humean approach. 

But there is really no reason that the sort of permissiveness offered by the best-systems account should not be extended to the realist setting, and so it appears there is a lacuna in the space of extant positions; it ought to be possible to provide a realist account of lawhood which can accommodate at least the same range of possible laws as the best systems account. The modal approach that we have suggested here is just  such an account. Indeed, it is straightforward to see that any putative law which can be written as the axiom of a `best-system' can also be written in the constraint framework. To see this, recall that Lewis' later accounts of the best-system approach distinguish two types of axioms\cite{Lewis1981}. Deterministic axioms simply state facts about the actual Humean mosaic, e.g. `everywhere in the mosaic a body's acceleration is equal to the net force on the body divided by its mass.' We can always rewrite such an axiom as a constraint according to the recipe `axiom $\rightarrow$ the set of all Humean mosaics in which the axiom is true.' Probabilistic axioms, meanwhile, \emph{`pertain not only to what happens in history, but also to what the chances are of various outcomes in various situations'} e.g. `atoms of type $A$ always have probability $x$ of decaying within the next $h$ seconds.' We can always rewrite such an axiom as a probability distribution over constraints according to the recipe `axiom assigning probability $p$ to some event $e$ $\rightarrow$ distribution assigning probability $p$ to the set of all mosaics in which event $e$ occurs,' or alternatively by using a nomic frequentist approach where `axiom assigning probability $p$ to some event $e$ ' is rewritten as the constraint `event $e$ must occur in a proportion $p$ of all cases of this kind across all of spacetime.' 

Thus the constraint framework can accommodate any type of law which the best-systems framework can accommodate, which means that it is of value to the ongoing debate over realism vs Humeanism - for if we are to make a fair comparison between realist and Humean approaches to lawhood, we should be comparing the versions of the two approaches which are the best fit to the reality of modern physics as we currently understand it. Thinking about laws in terms of constraints brings realist analyses of lawhood up to date with the sorts of developments discussed in section \ref{intro} and thus makes it clear that at least one of Humeanism's apparent advantages over realism - its ability to accommodate the large variety of laws postulated in modern science - is not a real advantage, as an appropriate realist analysis can achieve similar flexibility. Clearing away these sorts of considerations should allow the debate to be more clearly focused on the central epistemic and metaphysical distinctions which separate the realist and Humean pictures. 

 Is the converse true - can every probability distribution over constraints be written as an axiom of a best-system? Well, to start with, we saw in section \ref{Hume} that the best systems approach has trouble with laws which deal with   \emph{possibility} and \emph{impossibility} - in particular, it's hard for the Humean to make sense of laws like `X is possible' within worlds in which X does not actually occur. On the other hand the constraint framework certainly has the resources to accommodate these sorts of laws because it is structured around laws which delimit what is and is not possible.  For example, a law of the form `X is impossible' will straightforwardly induce the constraint consisting of  the set of mosaics in which X does not occur. Note that this does not entail that  whenever X does not occur  in the actual mosaic it is a law that X is impossible:  in such a case the actual mosaic does indeed belong to the set of mosaics in which X does not occur, but  the constraint framework tells us that not every set of mosaics to which the actual mosaic belongs to is associated with a law, because lawhood is a matter of objective fact which is not simply read off the mosaic. Laws of the form  `X is possible,' on the other hand, will \emph{not} typically be understood as inducing the constraint consisting of the set of mosaics in which X does occur, because we have seen that sometimes it is important to be able to insist on the lawlike nature of `X is possible' even if X does not occur in some possible world. Thus it is preferable to analyse a law of the form `X is possible' as the \emph{absence} of a law of the form `X is impossible,' so in the constraint framework, `X is possible'  is analysed as `there is no law inducing the constraint consisting of the set of all mosaics in which X does not occur,' which of course may be the case even if it so happens that the actual mosaic is a mosaic in which X does not occur. Thus the constraint approach is able to provide meaningful analyses of laws like `X is possible' even in cases where the best-systems approach will struggle. 
 
 Moreover, it is built into the best-systems approach that the actual laws of nature must be `simple,' whereas there are many possible constraints which would not lead to  axioms which look `simple' according to any reasonable description. We have made no requirement in our framework that the actual laws of nature must induce constraints which are simple; so the constraint framework  allows us to consider laws which would not be simple enough to be taken seriously within the best-systems framework.   This highlights an important distinction between   realist and non-realist approaches to lawhood. For the realist, the laws of nature are an objective fact about reality and whether or not they are simple is something which we might hope to discover by observation. Of course, in practice we tend to assume that the laws of nature are reasonably simple, since otherwise the problem of discovering them becomes intractable, but in the realist picture this simplicity principle is a \emph{hypothesis} rather than a matter of definition. Whereas for the Humean, the laws of nature are by definition whatever axioms lead to the simplest possible description, so simplicity is non-optional. And this has practical consequences: for the realist, simplicity is just one of many hypotheses which we may adopt to narrow down the search space for possible laws, so it could be partially or completely discarded if the empirical evidence gave us reason to think the laws are in fact not simple in some particular case, whereas for the Humean simplicity is part of what it is to be a law and so the simplicity constraint can never be abandoned.

  This feature may be of particular relevance as parts of physics begin to move away from the time-evolution paradigm, because once we start dealing with other sorts of laws it's possible that we will encounter laws that look more or less simple from different points of view. For example, we can certainly imagine laws applying to the whole of history all-at-once in an atemporal fashion which look simple from the external point of view but which look very complicated if we try to write down their  consequences for human observers (in much the same way as the motion of the planets looks mathematically simple if we consider the sun to be at rest but very complicated if we take the earth to be at rest). It's unclear whose point of view the axioms of the best system are supposed to be simple from - the original formulation in terms of a `gods-eye' description of the whole of history\cite{Dorst2019-DORTAB-2} does appear to suggest that simplicity from the external point of view is the key feature, but later strands of Humean thought such as the `better best systems approach'\cite{Cohen2009-COHABB} have emphasized the importance of coming up with laws which are simple and usable from the point of view of human observers. This ambiguity can mostly be swept under the carpet when we are dealing with time-evolution laws, which don't look notably different from the external and local points of view, but as physicists move beyond the time evolution paradigm it seems likely that this will become an increasingly pressing question. Thus a particularly interesting feature of the constraint framework is that it provides a new locus for these sorts of hypotheses. For example, rather than supposing that laws will look simple when expressed as dynamical equations, what if we instead hypothesize that they are simple in the sense that it should be possible to demarcate the relevant constraints in some way that is more efficient than simply specifying all the mosaics which are in the set? That is to say, perhaps we should be aiming to come up with laws which offer a simple description of the universe as a whole, rather than focusing on  the simplicity of their description of moment-by-moment dynamical evolution. We know from experience that assumptions like simplicity often work reasonably well in science, but there is no reason to assume that the specific way in which we have applied these assumptions in the past is optimal: applying old assumptions in a global way rather than  a local way may well open up new routes for scientific advancement.

\subsection{ Probability and Determinism \label{probability}}

The constraint framework allows us to disambiguate several different ways in which laws may be deterministic. The traditional notion of determinism pertains to what is often referred to as
\emph{Laplacean determinism}, where the result of a measurement at a given time is wholly determined by the state of the world at that time\cite{laplace1820theorie}. This definition makes sense in the time evolution picture; but in the context of non-dynamical laws there's no particular reason to think that `the state at a time' has any special significance, and so we are in need of a new generalized notion of determinism, which we may arrive at using our new generalized notion of lawhood. 

For example, we might choose to say we have  determinism at a `global' level if all of the laws of nature induce trivial distributions i.e. they only assign probabilities of $1$ and $0$. In that case we have no need for any objective chances, and each law is simply associated with a single constraint. Note that this sort of global determinism does not entail Laplacean determinism - for example, a law of the form `a certain process $P$ produces outcome $A$ with probability 0.8 and outcome $B$ with probability 0.2,' could be analysed as a constraint to the effect  `across all occurrences of process $P$ over all of spacetime, 80 percent have outcome $A$ and 20 percent have outcome $B$,' so process $P$ would therefore look indeterministic from a local point of view, but from the global point of view there would be no objective chances involved. So the constraint-based approach makes it clear that even though our best current theories don't seem to obey Laplacean determinism, it could certainly still be the case that we have something like determinism at a global level.

In the case where we have global determinism and the intersection of the corresponding constraints contains only a single mosaic, then the actual Humean mosaic is singled out uniquely by the laws of nature, so we have what one might describe as \emph{strong global determinism}. If the probability distributions induced by the laws are all trivial but the intersection of the corresponding constraints contains more than one mosaic, then the actual Humean mosaic must be selected from the intersection, so we have what one might describe as \emph{weak global determinism}. The latter can still be regarded as a form of determinism because the selection of a mosaic from the intersection is not a `chancy' event: the laws do not define any particular probability distribution over the mosaics in the intersection, not even the uniform distribution, since any one of them will do just as well as any other. The selection of the actual Humean mosaic from this set is therefore analogous to the selection of the initial conditions in in Laplacean determinism: it is arbitrary and it is not determined by anything, but it is also not chancy. Weak global determinism is thus simply a generalisation of the idea that in Laplacean determinism, both the initial conditions and the laws are needed to determine a unique course of history. 

Observe that the constraint framework provides us with a straightforward way to articulate the differences between Laplacean determinism, weak global determinism and strong global determinism - these distinctions could certainly not be formulated in the language of dynamical time evolution laws. Note further that it is important for this analysis that our approach is a \emph{realist} one, since if laws do not have modal force there can be no fact of the matter about whether they single the actual mosaic out uniquely, so a Humean analysis would necessarily look quite different. So the constraint framework does indeed seem to be fulfilling its intended purpose by providing a means of re-assessing topics like determinism  in the context of realist but non-dynamical laws.

\subsection{Unification} 

One important consequence of the generality of the constraint approach is that it allows us to express very different sorts of laws within the same framework. This is important for several reasons. First, if we want to perform analyses of philosophical ideas drawing on the notion of lawhood without leaning too strongly on specific assumptions about the form of the laws in question, it's necessary to do so using a framework which can accommodate a wide variety of laws. For example, traditional definitions of determinism intended for the time evolution picture don't work well for non time evolution laws, but below we will suggest a way to define determinism in a way that makes sense for any set of laws which can be expressed in the constraint framework. Thus the constraint approach allows us to make meaningful comparisons between the properties of very different sorts of laws.

Second, we may sometimes want to consider a case where we have very different sorts of laws acting within one and the same world. Prima facie it may seem difficult to understand how a set of very heterogenous laws could be applied together: for example, how do we combine a constructor-type law  with a time evolution law? Which law do we apply first in any given situation? Which is the more fundamental? But the constraint framework provides a straightforward way to combine different sorts of laws - and as long as there is at least one mosaic in the intersection of all of their constraints we can be assured that we have no contradictions between them. By contrast, the Humean approach can in principle accommodate all these sorts of laws, but it may be less clear whether we can make sense of a a wide variety of types of law all occurring within a single world, because one would typically expect that the axioms of the `best deductive system' would all be of a roughly similar type. For example, in any reasonable deductive system it should always be clear how to combine any set of two or more different axioms, and one might worry that a deductive system which combines both constructor-type laws and time evolution laws would not necessarily have that feature. Thus in fact the constraint framework  suggested here might actually be useful to Humeans too, since a Humean wanting to postulate a set of highly heterogenous laws could always use the constraint language as a way of formulating their axioms, and simply omit the supposition that constraints represent objective modal structure.  

And third, we noted in section \ref{intro} that physicists and philosophers are increasingly questioning the meaningfulness of the traditional distinction between kinematics and dynamics. Moreoever, we saw in section \ref{mosaic} that the requirements we would traditionally refer to as `kinematics' can be written in the form of a constraint consisting of all and only those Humean mosaics which contain only entities of the sort featured in the relevant kinematical space. We can then impose further constraints singling out smaller subsets of that set of Humean mosaics, and those further constraints play the role of the dynamics. So the constraint framework allows us to express both kinematics and dynamics in the same formalism, and thus it represents an important step towards delivering on the vision discussed in section \ref{intro}

\subsection{Higher Modal Structure} 

The framework we have suggested here has the further advantage that once we have made some proposal about the laws of nature using the constraint language, we can easily draw conclusions about the `higher-up' modal structure induced by these laws of nature between various less fundamental features of reality.

For example, consider two constraints, $A$ and $B$, such that $B$ is a proper subset of $A$ - that is to say, every Humean mosaic which satisfies the condition associated with $A$ must also satisfy the condition associated with $B$. Equivalently, we can say that every possible world containing a Humean mosaic satisfying the condition associated with $A$ must also be a possible world containing a Humean mosaic satisfying the condition associated with $B$. This of course is just another instance of metaphysical necessitation\cite{https://doi.org/10.1002/malq.19630090502}. Thus the fact that $B$ is a proper subset of $A$ can be understood as expressing the fact that $A$ metaphysically necessitates $B$, so the constraint framework provides a convenient way of studying the modal structure associated with relations between different sorts of regularities.  For example, $A$ could be the constraint associated with the no-signalling principle and $B$ could be the constraint associated with the no-signalling monogamy bound: it follows from the proof given in ref \cite{Toner} that $B$ is a proper subset of $A$, and thus we are licensed to say that the no-signalling principle metaphysically necessitates the no-signalling monogamy bound.  

Moreover, the same argument can be made in the case where $A$ is associated with a law whilst $B$ is a constraint which merely expresses a contingent fact - for example, $A$ could be the set of all mosaics in which it is everywhere true that $F = ma$ while $B$ could be the set of all mosaics in which it is true that on some particular occasion $F = ma$. Clearly here it will  be the case that $B$ is a proper subset of $A$ and hence in this setting the relation of metaphysical necessitation also underlies the relations between laws and individual instances of laws, allowing us to unify relations between regularities, and relations between laws and their instances, in terms of a single overarching form of modal structure.

\subsection{The Past Hypothesis} 

One argument that has been put forward in favour of the best systems approach is that it allows us to regard the past hypothesis as a law of nature rather than a mere coincidence\cite{Callender2004-CALTIN-2}. Now, realists about laws might well feel that the supposed gain here is a spurious one, for the best systems account has made the past hypothesis nomological only in virtue of making the nomological into a brute fact: in the Humean picture laws are just descriptions of whatever actually happens, so the fact that the past hypothesis is a law doesn't make it any less mysterious, and thus insofar as one finds the past hypothesis suspiciously coincidental from the realist point of view, one should also find it suspiciously coincidental within the Humean picture.

However, there is nonetheless an important insight here. The reason the past hypothesis is not normally treated as nomological is that physicists typically expect that physical systems can be decomposed into kinematics and dynamics; the initial state of a system is an element of the kinematics and is therefore by definition not a law, since laws are supposed to be dynamical. But as Hicks points out, Humeanism in its standard form does not accommodate any such distinction\cite{10.1093/bjps/axx006}: if including the initial state in the best-system leads to a significant simplification, it follows on the Humean view that the initial state is a law. Hicks regards this as a problem to be rectified, but from the present perspective it is a significant advantage: the Humean view gives us licence to ignore arbitrary distinctions such as the kinematics/dynamics split and simply write down whichever laws provide the best systematization, regardless of whether they accord with common assumptions about what laws ought to look like. 

So really the novel contribution of Humeanism to the problem of the past hypothesis is located not in its antirealism but rather in the fact that it licences us to take nonstandard, temporally non-local laws seriously, and thus allows us to put the initial state into the laws of nature. But we do not have to become Humeans in order to achieve this end - we simply have to reject the standard kinematics/dynamics distinction. The constraint framework straightforwardly  allows us to include the past hypothesis in the laws of nature, since it can be understood as a constraint which assigns probability $1$ (or at least, high probability) to the set of Humean mosaics where the initial state is a `low entropy' state (or some more sophisticated description\cite{pittphilsci8894}).

So the constraint framework allows us to make the past hypothesis nomological in a more robust sense than the Humean approach: it is not merely a fact about the universe which happens to appear in the best-system, but it is determined by some law that is part of the objective modal structure of reality and thus it can be given a lawlike explanation in just the same way as anything else in the universe.

\section{Objections} 

In this section, we discuss some possible objections that might be raised to the analysis of lawhood we have presented 

\subsection{Epistemic Issues} 

If constraints are ontologically prior to the Humean mosaic, then it seems that laws of nature are necessarily external to the Humean mosaic, and this feature forms the basis of a common objection to realist accounts of lawhood: if laws are not a part of the Humean mosaic, and we are only able to observe directly those things which feature in the Humean mosaic, how can we ever come to know about or have grounds for believing in laws of nature? To address this concern, it will be important to distinguish between  two different epistemic difficulties. First, how could we ever know or have grounds to believe that laws which are ontologically prior to the Humean mosaic exist? And second, how could we find out about their true metaphysical nature?

With regard to the first difficulty, the answer seems straightforward: we infer that laws exist based on our observations of their effects on the world. That is to say, we observe regularities and then apply inference to the best explanation to conclude that these regularities are the effect of some objective structure, which is either the consequence of or to be directly identified with the laws of nature. For as Hildebrand argues, if an entity is external to the actual universe but is essential to the analysis of lawhood, then in fact we do have a way to find out about it - to wit, by studying the laws of nature\cite{Hildebrand2020-HILPLO}. Here, we are specifically identifying laws of nature as those things which are external to the physical universe but which \emph{have an effect on it}: that is the content of the claim of ontological priority. So if there are any laws which satisfy this criterion, they are certainly not disconnected from physical reality - they act on physical reality, even if physical reality cannot act back on them, so we do have grounds for forming beliefs about them.

With regard to the second difficulty, as we have adopted a structural realist approach we should simply hold that we can be epistemically committed to the real modal structure associated with laws without knowing anything about the underlying metaphysical nature of the laws (whether because we have no epistemic access to laws in and of themselves, in the tradition of ESR, or because laws \emph{are} nothing over and above modal structure, in the tradition of OSR). Of course, this approach might rule out some uses to which one might want to put one's realism about lawhood: if for example  one were attempting to use an analysis of lawhood to solve the problem of induction, then indeed one would potentially need to answer questions about how laws actually operate on physical reality, in order to make the argument that the existence of laws guarantees that the future will be relevantly similar to the past. But if one is a realist simply because the existence of governing laws seems like the best explanation for observed regularities, or because postulating laws leads to a more coherent picture of the content of reality, then there is no obvious need to explain \emph{how} laws bring about the effects that they do.

A more specific variant of this objection to realism involves the concern that it is hard to see how universals or capacities or any such metaphysical notion could play the role in rational belief-formation that laws clearly do. This is sometimes known as the inference problem - that is, the problem of giving an account of how it is that laws permit us to make inferences\cite{Hochberg1981-HOCNNA, Fraassen1993-FRAACA-2, university1987philosophical}. It seems likely that a similar complaint would be levelled at our more general claim that laws induce constraints which are ontologically prior to the Humean mosaic - why, the Humeans will ask, should we base our expectations about future events on something that is not even a part of the physical universe? But again, t the claim that the inference problem can only be answered by giving an account of the metaphysical nature of the laws should be resisted. According to the view we have put forward here, laws are by definition those things which induce certain modal structures, or perhaps laws simply \emph{are} modal structures, so if we believe that certain laws hold then we have every right to form expectations about the effects that they will bring about, since that is exactly what the relevant modal terminology is designed to capture. In much the same way, we can make inferences about the future based on our knowledge of electrons because even if we don't know what electrons actually are we do have a clear understanding of them in structural and functional terms: electrons are those things which behave in certain ways, so if we believe that electrons do exist and are present in a given instance, we can make inferences about the effects they will produce.

Of course, one might attempt to make a distinction on the grounds that in the case of electrons there is at least the possibility in principle of finding out more about the nature of the underlying constituents, whereas in the case of laws it is most likely impossible even in principle for us to find out the true metaphysical nature of the laws. Is this an epistemically relevant distinction? Those who consider themselves scientific realists but who also espouse Humean (or other non-realist) approaches to lawhood certainly seem to think so - the claim seems to be that there is something particularly problematic about making inferences to things which are unobservable because they are external to the Humean mosaic, whereas it is less problematic to make inferences to things which are unobservable for some more prosaic reason. But this particular way of setting the boundary of the knowable seems fairly arbitrary. Indeed, we are never going to observe the metaphysical basis for laws `directly,' if  there is any such basis at all, but then we are almost certainly never going to observe electrons `directly' either, since even in a cloud chamber what we actually observe are the effects of the electron rather than the electron itself. And if we have learned anything from the long history of debate over the distinction between observable and unobservable objects, it is that if we countenance only beliefs in directly observable features of physical reality then very little is safe. Realism about laws is in this in this regard not importantly different from other forms of scientific or metaphysical realism: in either case we have solid grounds for realism provided we have a solid understanding of the modal properties of the entity in question. 

A slightly different objection regards the role of laws in explanation. It has been argued that explaining regularities by appeal to laws does little to remove the mystery surrounding them, because we have just swapped the question `Why are there regularities in the universe?' with the equally puzzling `Why should laws give rise to regularities?' And the same point can be made about the account in terms of constraints - there are many possible constraints which would not give rise to any noticeable regularities, and so appealing to constraints does nothing to explain the existence of regularities without some further supposition about why the laws of nature happen to induce the special sorts of constraints which produce noticeable regularities. We have not attempted to make such a supposition because we see no way to do so without invoking some kind of hypothesis about the underlying metaphysical nature of the laws, and the point of focusing on structures was precisely to avoid those sorts of commitments. Nor is such a supposition necessary, because the account we have given is not intended to solve the explanatory problem: it is only supposed to offer a picture of the world which allows regularities to be the consequences of objective modal structure.  The question of \emph{why} there should be any objective modal structure, or structure of this particular kind, is another matter altogether.

\subsection{Law Selection}

Going against the thrust of our approach in this paper, one might argue that it is actually a good thing that traditional realist approaches based on a time evolution picture severely constrain the space of possible laws, because otherwise there will simply be too many possible laws and we will never be able to get the underdetermination down to a reasonable level. Indeed, it's true that scientists often need to constrain the space of possible laws in some way in order to make the search space tractable. That is the purpose served by criteria such as `simplicity,' `unification,' `beauty,' `locality,' `unitarity,' and so on. And it's also true that insisting that laws should take the time evolution form (or at least, that it should be possible to rewrite them in a time evolution form) is one way in which scientists have historically constrained the search space. 

But scientists also recognise that sometimes constraints on the search space need to be abandoned when the empirical data demonstrates their unsuitability. For example, in the aftermath of Bell's theorem\cite{Bell} many physicists came to the conclusion that it was necessary to abandon the principle of locality\footnote{I reinforce that `locality' as referenced in Bell's theorem is not the same as the mathematical principle of `locality' employed in quantum field theory; Bell's theorem says nothing about the latter.}\cite{Seevinck} which was previously regarded as a natural constraint on the set of possible laws. And as we saw in section \ref{intro} a variety of physicists have now begun exploring laws which move away from the time evolution picture, driven in many cases by the conviction that the empirical data have shown the time evolution form to be unsuitable. Of course this is not yet a consensus position, and it remains possible that science will find a way to accommodate all the existing data within a strict time evolution model, but given the increasing prominence of these approaches it seems the right time to begin exploring some of their consequences for the philosophy of lawhood.

\subsection{Cartwright's Challenge} 

The constraint methodology works straightforwardly for certain very simple laws - for example, `no signal can propagate faster than light.' Subject to a satisfactory definition of what counts as a signal, if this is a law it is obeyed exactly, and  therefore it can very straightforwardly be translated to a constraint consisting of the set of Humean mosaics which do not contain any signalling processes.

But what about a law such as the Newtonian gravity relation, $F = \frac{GMm}{r^2}$? We cannot simply identify this law with the set of all mosaics in which it is everywhere the case that  $F = \frac{GMm}{r^2}$, because a `force' is not the sort of thing which is typically supposed to feature in a Humean mosaic. We would have to combine the Newtonian gravity law with the force law $F = ma$; but we also cannot identify this law with the set of all mosaics in which it is everywhere the case that  $ma= \frac{GMm}{r^2}$, because as Cartwright famously pointed out,\cite{Cartwright} this relationship is never exactly satisfied, since bodies are typically subject to more than one force and their final acceleration is given by the vector sum of all the forces acting on them. Moreover, we also cannot make $ma= \frac{GMm}{r^2}$ into a probabilistic law, because although bodies do not usually obey  $ma= \frac{GMm}{r^2}$ exactly, it is also not the case that their motion is probabilistic - it is a deterministic function of all the forces acting on them. 

One possible resolution of this problem would be to identify a constraint $C$ with the set of worlds for which $C$ belongs to the axioms of the best-system, defined in the usual Humean way as the system which optimizes simplicity, strength and fit. But the problem with this approach is that it is subject to the same objection often levelled against Humeanism - i.e. that `simplicity' is not objective and that therefore what counts as a law will depend on our preferred standards of simplicity\cite{10.2307/43154224}. This is a consequence that the Humean might be willing to accept, but it is significantly more problematic in our case, since we have required that laws should be ontologically prior to the Humean mosaic and so it seems we cannot allow that they should depend on the preferences of the inhabitants of the mosaic. 

An alternative would be to reconsider the idea that forces should not feature in the Humean mosaic; then we might be able to identify a set of mosaics in which it is exactly true everywhere that there exists a force $F = \frac{GMm}{r^2}$, and also exactly true that the resulting motion is the vector sum of the forces. One's willingness to accept this solution will naturally be dependent on one's intuitions about whether a `force' counts as a local matter of particular fact. We think it is unlikely that Lewis would have liked this, but we do not necessarily have to follow Lewis on this point. Moreover within the constraint picture we have a little more leeway, because we have already abandoned strict Humean supervenience and therefore we need not be held back by the fear of betraying our founding metaphysical principles. 

A final possibility would involve abandoning the Newtonian gravitational law in favour of conservation laws or a Lagrangian description. For while it is not exactly true that $ma = \frac{GMm}{r^2}$ anywhere, it  \emph{is} exactly true that momentum, energy and angular momentum are conserved for a closed system, and from these conservation laws it is possible to deduce the individual accelerations of the constituent systems. Of course in the real world very few systems are truly closed, but this is not a  problem for the constraint view, because constraints are in general non-dynamical, so we can make the system as large as necessary to ensure that the conservation is exact - even going so far as to include the whole universe if need be. 
Indeed, one might argue that the reason Newtonian force laws are very seldom exactly true is that they are not the `real' laws - it is the conservation laws or Lagrangian descriptions which feature in the true laws of nature, and force laws are simply the result of our attempts to parse the messy local consequences of these global laws. From this point of view, Cartwright's challenge results from supposing that laws must be spatially and temporally local, whereas in fact laws are generically non-local and therefore hold exactly only at a global level. 

Of course, it may be objected that although this works for the case of Newtonian force laws because we always have the alternative Lagrangian description available to us, there are conceivable laws where this sort of description is not possible - indeed, it is known that a set of dynamical laws has a corresponding Lagrangian description if and only if, after transformation into a certain canonical form, the right-hand sides of all the equations are derivable by differentiation from a single function $H$\cite{lanczos, Ostrogradsky, Butterfieldaction}. And even if it is in fact the case that all laws in the actual world do have a Lagrangian equivalent, the stated intention for the constraint framework was to come up with a highly general language which avoids ruling out large classes of laws by fiat, so it would be a problem if  this framework can't deal with laws which don't have a Lagrangian description. 

However, for any set of deterministic laws which are both individually and jointly deterministic, there must be some way of writing these laws (perhaps in the form of a single highly complicated disjunctive law, and/or by augmentation with a suitable composition rule) such that they hold exactly. If this were not the case, then there would be some scenarios in which the set of laws did not prescribe a single outcome, meaning that the laws would not after all be jointly deterministic. For example, in the force law case, even if the Lagrangian description were not available it would be possible to find a law which is exactly true by writing $ma = \int F(x)$ where the integral (or sum in the finite case) is taken over all sources of force. The resulting law might be highly non-local, extremely complicated and aesthetically displeasing, but nonetheless we would be able to pick out a set of Humean mosaics in which this law always holds and thus it could be written as a constraint. And a similar argument  shows that any set of laws which individually and jointly prescribe exact probabilities can be written as an exact disjunctive law assigning probabilities over some disjunctive sample space of events, meaning that together they can be written as a probability distribution over constraints.

\subsection{The God Problem} 

It is common to criticize realist accounts of laws of nature on the grounds that they get their plausibility from implicit (or sometimes explicit!) invocation of a deity who personally steps in to ensure that the universe obeys the laws of nature; thus it is argued that such accounts are a remnant of theism and should be rejected by all right-thinking atheists\cite{Orr2019-ORRNGN, CartwrightManuscript-CARNGN}. And one might make a similar criticism of the approach we have suggested here, for although we have steered clear of making any assertions about the metaphysical underpinnings of the laws of nature, nonetheless there seem to be some vaguely theistic ideas in play. In particular, in the discussion of constraints we invoked the notion of `ontological priority,' which is  often understood in terms of the order in which God creates things, and thus  one way to motivate the idea of the Humean mosaic being subject to constraints would be to picture the universe as something like a sudoku grid subject to global rules, with God stepping in to solve it. 

But in fact we should take care with this metaphor, because it seems to be based on an intuition that the parts of the universe must come into being in some temporal order, with God's process of solving the puzzle constituting a process of temporal becoming. Yet the universe, considered as an atemporal whole, already contains within it the temporal relations which are constitutive of our experience of time; and therefore one might argue, a la McTaggart, that we do not need an external `A-Series' process of temporal becoming in addition to the existing `B-Series' determined by the temporal relations internal to the universe\cite{McTaggart}. Thus it seems that the metaphor which conjures up God solving a puzzle adds an unnecessary layer of temporal becoming to an account of lawhood which is perfectly adequate without it.  

And indeed, it is easy to remove God from the picture. For after all, a sudoku grid does not actually require a player: it has a unique correct solution regardless of whether anyone ever comes along to solve the puzzle. We need only suppose that the universe is the same: the constraints prescribe (either deterministically or probabilistically) a set of possible solutions, and the actual Humean mosaic exists and satisfies these constraints in an atemporal block universe sense, with no need for a personal being to come along and solve the puzzle. 

Of course, in the context of \emph{epistemic} structural realism about laws it is part of the deal that we can't know what laws of nature are or how they get enforced, so although ontic structural realists might be able to be dogmatic about this, epistemic structural realists can't - for all they know, perhaps there is a God out there solving the puzzle of reality, and/or the universe really is generated in a process of temporal becoming. But the point to be made here is that the plausibility of the approach does not depend in any way on the theistic metaphor, and indeed the theistic metaphor may well be more harmful then helpful.

\section{Conclusion} 

Developments in modern physics give us good reason to take seriously the possibility of laws which are non-local, global, atemporal, retrocausal, and/or not easily put in the standard kinematical-dynamical form. We have seen that existing realist accounts of laws as relations between universals or capacities/dispositions/powers are not well suited to accommodate these sorts of non-standard laws, and thus there is a clear need for a realist account which brings realism about laws up to date with developments in modern physics. Ideally, this account will bring some conceptual clarity to the new approaches currently emerging within physics, and will also give us the resources to re-assess some common philosophical applications of the concept of lawhood in the light of these scientific development. 

We. argued that this account can be given by appealing to the idea that laws are part of the objective modal structure of reality. However, existing analyses of modal structure in terms of individual systems and/or causal relations are not sufficiently general for this purpose, and thus we set out to provide a more general account of modal structure. Drawing on a number of physics examples, We proposed characterising laws in terms of probability distributions over constraints, and demonstrated that this approach is capable of accommodating a wide variety of possible laws. Finally, we gave some brief examples of possible applications of this framework to the philosophical literature on determinism, modal structure and the past hypothesis.

The examples we have offered here are intended only as a preview of the ways in which this framework might be applied; all of the discussion in section \ref{applications} could be significantly expanded, and we  have not yet touched on the status of laws in accounts of causation, explanation, prediction and confirmation. Moreover, once we take seriously the possibility of non-dynamical laws, there are important scientific questions about how researchers might furnish evidence for non-standard laws which can't straightforwardly be tested in ordinary laboratory experiments. My hope is that the constraint approach will furnish a clear and precise language in which some of these questions can be articulated and answered.

\section{Note}

Shortly before publication the author became aware of a similar paper which also analyses lawhood in terms of constraints, though within a more explicitly primitivist framework: see ref \cite{chen2021governing}. There is no connection between these two papers, but it is interesting to see convergence on these ideas from different sources.

\bibliographystyle{unsrt}
\bibliography{newlibrary11}{} 

\end{document}